# STRUCTURAL MORPHING METASURFACES FOR ELECTROMAGNETIC BEAM MANIPULATION

Aakash Bansal, William G. Whittow
*Wolfson School of Mechanical, Electrical, and Manufacturing Engineering, Loughborough University*
*Loughborough, United Kingdom*
a.bansal@lboro.ac.uk, w.g.whittow@lboro.ac.uk

*Abstract* – The paper presents a novel concept of a 3D structural metasurface with mechanically morphing capability that can be used as a transmit- or reflect-array to manipulate electromagnetic beams for applications in RF sensing and communications for both space and ground stations. They can be controlled using low-power actuators to deform the surface profile which is then utilised as a lens or a reflector for beamforming and steering. The proposed simulated structural metasurfaces can be used for either beam-steering or to generate bespoke contour beams for satellite communication and sensing.

## I. INTRODUCTION

Conventional intelligent surfaces commonly referred as reconfigurable intelligent surfaces or intelligent reflecting surfaces or metasurfaces are a popular candidate for applications in electromagnetic (EM) sensing and communications for both terrestrial and non-terrestrial networks [1]–[4]. They are composed of a large array of unit-cells (also known as meta-atoms) reconfigured using active components such as p-i-n diodes and varactors. They are used to manipulate the EM beam to control its characteristics such as polarization or direction [5], [6].

Several intelligent surfaces in the literature utilize thousands of such active components to control the incident beam and are being envisioned as a replacement for large dish antennas on a satellite. This allows us to control the radio environment efficiently, but it comes at the cost of high-power consumption, maintenance, and fabrication costs. Such metasurfaces need consistent power to maintain the beam reconfiguration and hence, are power hungry [7], [8].

As an alternative to such power hungry metasurfaces and weight-intensive reflector dishes, this paper presents a new shape-morphing metasurface architecture that can be used to optimally shape the surface structure and control the reflected beam through them either to generate a single highly directional beam or a multi-beam layout to generate different contours.

The concept of morphing metasurfaces has recently been seen in the literature but with very limited capabilities [9], [10]. Such structures can be used to create a more continuous beam-steering compared to traditional beam-switching techniques that are commonly seen in the literature [11]–[13], making them highly suitable for both terrestrial and non-terrestrial applications.

The rest of the paper is divided as follows: Section II provides an explanation of how to generate shape morphing ability within a metasurface; Section III offers simulated results for a shape-morphing reflector antenna followed by a conclusion in Section IV.

## II. SHAPE-MORPHING SURFACE

The shape-morphing for the proposed metasurface is derived from the 3D-printed structural techniques described in [14]. The 3D lattice of the structure allows the two-dimensional surface to mould in different shapes, which is then used to act as a reflective metasurface. An example of such morphing is shown in Figure 1.

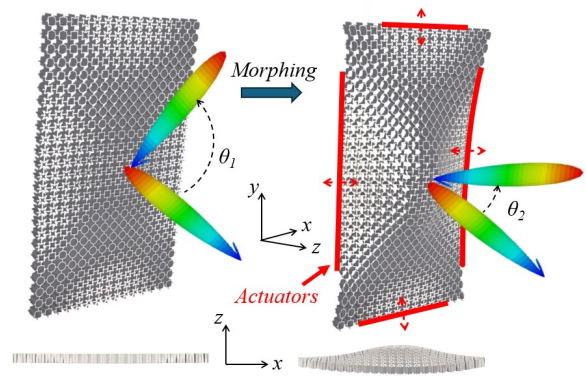

*Figure 1: A perspective view of the morphing of the 2D metasurface to create different 3D profiles and achieve beam-reconfigurability.*

The proposed design is a functionally graded structure designed for 3D-printed fabrication process and optimised to acquire any prescribed shape with force applied at different points. In [10], such a surface was used to generate a beam-scanning highly directive radio frequency (RF) dielectric lens. Here, with the inclusion of metal sheet, the beam is reflected back. The reflected beam properties are manipulated with the shape achieved by the surface.

The metal sheet shape is defined as a one-dimensional curve which is bent in a convex shape and is defined using Eq. (1),

$$y = \sin\left(\left(\frac{0.5}{r}\right)\cdot\left(\frac{\pi(x+x_s)^2}{r}\right)\right)\cdot\left(\frac{100}{r}\right).h \quad (1)$$

Here, $y$ is any arbitrary point on the cartesian plane, $h$ is the maximum height for the convex shape of the reflector metasurface, $x_s$ defines the shift in $x$-axis to redirect beam in required direction, and $r$ is the width of the metasurface. A 1D plot defined by Eq. (1) is presented in Figure 2. This keeps the length of the surface consistent, and by the way of applying force in different edges, the sheet is re-shaped to redirect the beam in different directions.

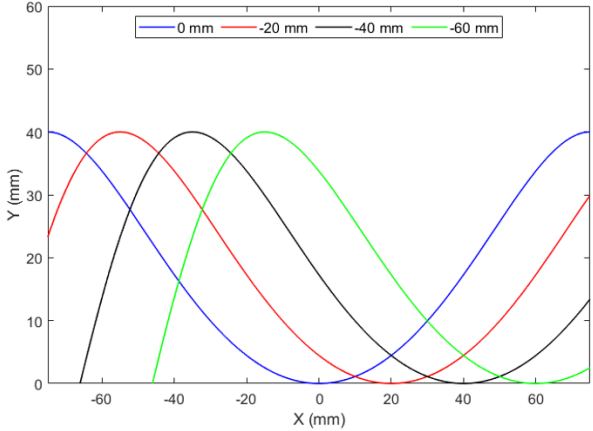

*Figure 2: One-dimensional profile of four different configurations of the morphing metasurface achieved by changing shift in x-axis, $x_s$.*

Within this study, we have presented four configurations of a morphing metasurface to redirect the reflected beam in azimuth plane. The simulation setup and four metasurface configurations are presented in Figure 3. The design was simulated using CST Microwave Studio, and the simulated results are discussed in Section III.

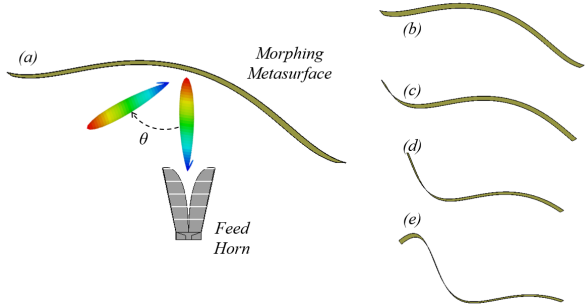

*Figure 3: (a) Top view of the morphing metasurface with the feed horn. The metasurface is tilted at an angle of 15° orthogonally from the feed horn. Different metasurface configs for shifts of (a) 0 mm, (b) -20 mm, (c) -40 mm, and (d) -60 mm, as seen in Figure 2.*

## III. SIMULATED RESULTS

A wideband ridged horn antenna design was used as a feed to the planar sheet, and simulation exercises were carried out for a frequency band of 8 to 15 GHz. The simulated radiation pattern at 10 GHz with and without the four configurations of the morphing metasurface are presented in Figure 4.

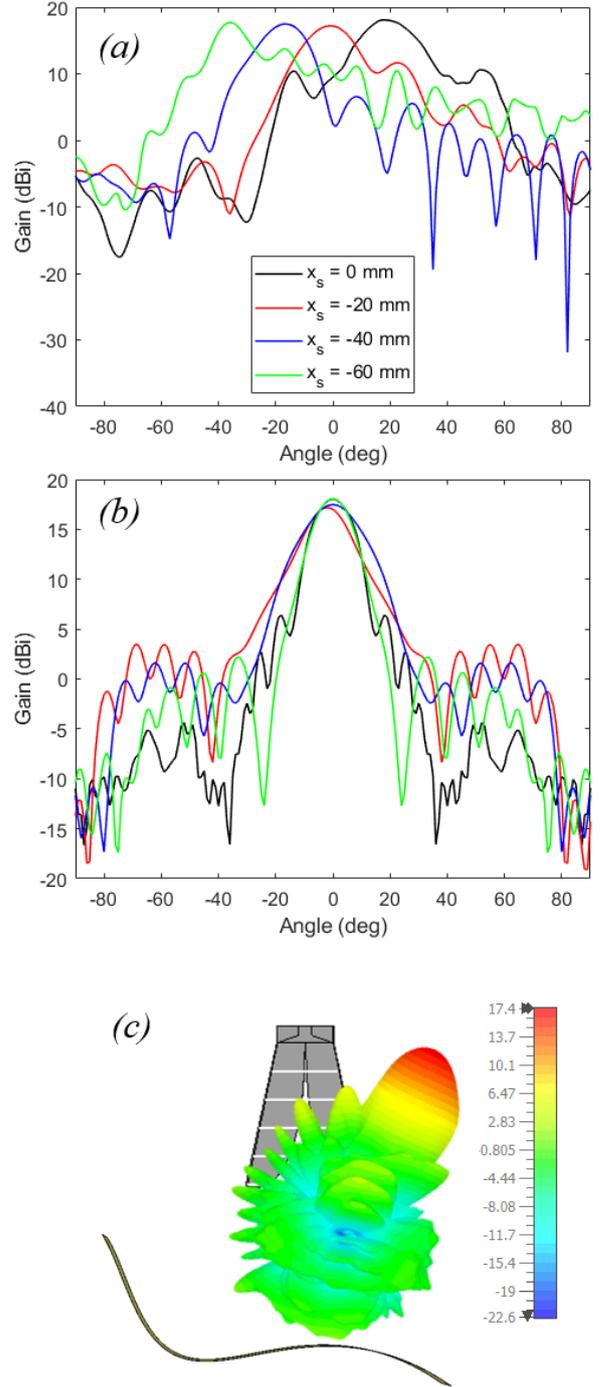

*Figure 4: Simulated 2D radiation pattern in (a) azimuth and (b) elevation planes for the four configurations of the metasurface. (c) Simulated 3D radiation pattern with third configuration ($x_s$ = -40 mm) showing the reflected beam.*

The antenna beam is steered between +36° and -18° with the shape morphing of the metasurface across the four configurations with $x_s$ ranging from 0 to -60 mm.

An 18⁰ beam steer was generated for every 20 mm shift in *x*-direction of the metasurface. The simulated gain of the antenna at 10 GHz for the four morphing configurations was found to be approximately 17.2 (± 0.3) dBi, hence, confirming the consistency in gain and performance of such morphing reflective metasurface. Similar performance was observed throughout the frequency band.

## IV. Conclusion

The article presented a novel very-low-power shape morphing reflective metasurface that can potentially replace large dishes and intelligent reconfigurable surfaces in both terrestrial and non-terrestrial telecommunication applications. The proposed morphing metasurface is an optimised functional structure that can be reshaped with boundary excitation using actuators. The proposed structure was simulated using CST Microwave Studio and offers a beam-steering of +36⁰ to -18⁰ with small shift in the reflective metasurface shape. The contraction and relaxation in the structure generated with mechanical stress can help generate out-of-plane deformations in different regions which can allow the beam not to just steer but also reshape to generate different contour beams.